\documentclass[prd,twocolumn,floats,floatfix,showpacs,nofootinbib]{revtex4}
\usepackage{graphicx}
\usepackage{dcolumn}
\usepackage{bm}
\usepackage{graphics}
\usepackage{slashed}
\usepackage{amssymb}
\usepackage{natbib}
\usepackage{amsmath}
\usepackage{amsthm}
\usepackage{url}
\usepackage{color}
\newcommand{\bea}{\begin{eqnarray}}
\newcommand{\ena}{\end{eqnarray}}
\newcommand{\bean}{\begin{eqnarray*}}
\newcommand{\enan}{\end{eqnarray*}}

\newcommand{\fracc}[2]{\frac{\textstyle{#1}}{\textstyle{#2}}}
\newtheorem*{theorem}{Conjecture}
\begin{document}

\title{Dragged metrics}

\author{M. Novello\footnote{M. Novello is Cesare Lattes ICRANet
Professor}}\email{novello@cbpf.br}
\author{E. Bittencourt}\email{eduhsb@cbpf.br}

\affiliation{Instituto de Cosmologia Relatividade Astrofisica ICRA -
CBPF\\ Rua Dr. Xavier Sigaud, 150, CEP 22290-180, Rio de Janeiro,
Brazil}

\date{\today}

\begin{abstract}
We show that the path of any accelerated body in an arbitrary
space-time geometry $ g_{\mu\nu} $ can be described as geodesics in
a dragged metric  $ \hat{q}_{\mu\nu} $ that depends only on the
background metric and on the motion of the body. Such procedure
allows the interpretation of all kind of non-gravitational forces as
modifications of the metric of space-time. This method of effective
elimination of the forces by a change of the metric of the
substratum can be understood as a generalization of the d\rq
Alembert principle applied to all relativistic processes.

\end{abstract}

 \maketitle

\section{Introduction}

In 1923 Gordon \cite{gordon} made a seminal suggestion to treat the
propagation of electromagnetic waves in a moving dielectric as a
modification of the metric structure of the background. He showed
that the waves propagate as geodesics not in the geometry $
\eta_{\mu\nu} $ but instead in the dragged metric
\begin{equation}
\widehat{q}^{\mu\nu} = \eta^{\mu\nu} + ( \epsilon \mu - 1 ) \,
v^{\mu} \,v^{\nu}, \label{251211}
\end{equation}
where $ \epsilon$ and $ \mu$ are constant parameters that
characterize the dielectric. Latter it was recognized that this
interpretation can be used to describe non-linear structures when $
\epsilon$ and $\mu $ depends on the intensity of the field
\cite{novello1}.

In recent years an intense activity concerning properties of
Riemannian  geometries similar to the one described by Gordon has
been done \cite{novello2}. In particular those that allows a
binomial form for both the metric and its inverse, that is its
covariant and the corresponding contravariant expressions
\begin{equation}
\widehat{q}^{\mu\nu} =  \eta^{\mu\nu} + b \, \Phi^{\mu\nu},
\label{2612111}
\end{equation}
and
\begin{equation}
\widehat{q}_{\mu\nu} = A \, \eta_{\mu\nu} + B \, \Phi_{\mu\nu}.
\label{2612112}
\end{equation}
Thus, the tensor $  \Phi^{\mu \nu} $ must satisfy the condition
\begin{equation}
 \Phi_{\mu \nu} \,   \Phi^{\nu \lambda} \, = m \,   \delta_{\mu}^{\lambda} + n \,   \Phi_{\mu}^{\lambda}.
 \label{2}
\end{equation}
In the present paper we limit our analysis only to the simplest
dragged form by setting $ \Phi^{\mu\nu} = v^{\mu} \,v^{\nu}.$ In
this case the coefficients of the covariant form of the metric are
given by

$$ A = 1; \hspace{0.8cm} B = - \frac{b}{1  + b}. $$

The origin of the dragging effect in the case of Gordon\rq s metric
is due to the modifications of the path of the electromagnetic waves
inside the moving dielectric. Then we face the question: could such
particular description of the electromagnetic waves in moving bodies
be generalized for other cases, independently of the electromagnetic
forces? In other words, could such geometrized paths be used to
describe other kinds of forces? We shall see that the answer is yes.
Indeed, we will show that it is possible to geometrize different
kinds of forces by the introduction of a dragged metric $
\widehat{q}_{\mu\nu}$ such that in this geometry the accelerated
body follows the free path of geodesics.

Let us emphasize that we deal here with any kind of force that has a
non-gravitational character. It is precisely the consequences of
such non-gravitational force that we describe in terms of a modified
dragged metric. This means that the observable effects of any force
can be interpreted as nothing but a modification of the geometry of
space-time. In other words the motion of any accelerated body can be
described as a free body following geodesics in a modified metric.
This procedures generalizes d\rq Alembert principle of classical
mechanics \cite{lanczos, mach} which states that it is possible to
transform a dynamical problem into a static one, where the body is
free of any force. Going from the background metric -- where an
accelerated body experiences a non-gravitational force -- to a
dragged metric where the body follows a geodesics and become free of
non-gravitational forces is the relativistic expression of this
principle. In this way we produce a geometric description of all
kind of motion whatever the force that originates it.

\section{A special case}

We claim that accelerated bodies in a flat Minkowski
space-time\footnote{Let us point out that the analysis we present in
this section may be straightforwardly generalized to arbitrary
curved Riemannian background. } can be equivalently described as
free bodies following geodesics in an associated dragged geometry.
In order to simplify our calculation we restrict this section to the
case in which the acceleration vector $ a_{\mu}$ is the gradient of
a function, that is\footnote{ Using the freedom in the definition of
the four-vector $ v_{\mu}$ we set $ v_{\mu} v^{\mu} = 1.$ The
acceleration is orthogonal to it, that is, $ a_{\mu} \, v^{\mu}
=0.$}

\begin{equation}
 a_{\mu} = \partial_{\mu}
\Psi. \label{31dez152} \end{equation} Thus the force acting on the
body under observation comes from a potential $ V $ in the
Lagrangian formalism, i.e., $ F_{\mu} = \partial_{\mu} V.$

We write the dragged metric in the form
$$  \widehat{q}^{\mu\nu} =  \eta^{\mu\nu} + b \, v^{\mu} \,v^{\nu}.
$$
The associated covariant derivative is defined by
 $$ v^{\alpha}{}_{; \, \mu} = v^{\alpha}{}_{, \, \mu} +
\widehat{\Gamma}^{\alpha}_{\mu\nu} \, v^{\nu}, $$
where the corresponding Christoffel symbol is given by
\begin{equation}
\widehat{\Gamma}^{\epsilon}_{\mu\nu} = \frac{1}{2} \, (
\eta^{\epsilon \alpha} + b \,v^{\epsilon} \, v^{\alpha} ) \,
(\widehat{q}_{\alpha\mu, \nu} + \widehat{q}_{\alpha\nu , \mu} -
\widehat{q}_{\mu\nu , \alpha} ), \label{261211520}
\end{equation}
where we are using a comma to denote simple derivative, that is $
A_{, \mu} \equiv \partial_{\mu} A .$  The description of an
accelerated curve in a flat space-time as a geodesics in a dragged
metric is possible if the following condition is satisfied
\begin{equation}
\left( v_{\mu , \nu} - \widehat{\Gamma}^{\epsilon}_{\mu\nu} \,
v_{\epsilon} \right) \, \widehat{v}^{\nu} = 0, \label{31dez1}
\end{equation}
where we have used the dragged metric to write $ \widehat{v}^{\mu}
\equiv \widehat{q}^{\mu\nu} \, v_{\nu}.$ Or, equivalently,
\begin{equation}
\left( v_{\mu , \nu} - \widehat{\Gamma}^{\epsilon}_{\mu\nu} \,
v_{\epsilon} \right) \, v^{\nu} = 0 \label{31dez1}.
\end{equation}

Then noting that the acceleration in the background is defined by  $
a_{\mu} = v_{\mu , \nu} \, v^{\nu} $ and using equation
(\ref{31dez152}) the condition of geodesics in the dragged geometry
takes the form

\begin{equation}
 \partial_{\mu} \Psi = \widehat{\Gamma}^{\epsilon}_{\mu\nu} \, v_{\epsilon} \,
v^{\nu}.
\end{equation} We have
\begin{equation}
\widehat{\Gamma}^{\epsilon}_{\mu\nu} \, v_{\epsilon} \, v^{\nu} =
\frac{1 + b}{2} \, v^{\alpha} \, v^{\nu} \, \widehat{q}_{\alpha\nu ,
\mu}. \label{31dez15}
\end{equation}
Using the expression of $ \widehat{q}_{\alpha\nu} $ and combining
with condition (\ref{31dez1}) it follows
$$ a_{\mu} + \frac{\partial_{\mu}b}{2 (1 + b)} = 0, $$ that is, the
expression of the coefficient $b$ of the dragged metric is given in
terms of the potential of the acceleration
\begin{equation}
1 + b = e^{- 2 \Psi}. \label{31dez155}
\end{equation}

\subsection{The curvature of the dragged metric}

In the case the background metric is not flat or if we use an
arbitrary coordinate system the connection is given by the sum of
the corresponding background one and a tensor, that is

\begin{equation}
\widehat{\Gamma}^{\epsilon}_{\mu\nu} = \Gamma^{\epsilon}_{\mu\nu}
 + K^{\epsilon}_{\mu\nu}.
\end{equation}
In the case of the Minkowski background a direct calculation gives
for the connection $ \widehat{\Gamma}^{\epsilon}_{\mu\nu} $ the form
\begin{equation}
K^{\epsilon}_{\mu\nu} = v^{\epsilon} \, ( a_{\mu}\, v_\nu + a_{\nu}
\, v_{\mu} ) - a^{\epsilon} \, v_{\mu} \, v_{\nu}. \label{31dez17}
\end{equation}
Then,
$$ K^{\epsilon}_{\mu\epsilon} = a_{\mu}.$$
The contracted Ricci curvature has the expression

\begin{equation}
\widehat{R}_{\mu\nu} = a_{\mu , \nu} - a_{\mu} \, a_{\nu}  + (
\omega + a^{\alpha}{}_{, \alpha} ) \, v_{\mu} \, v_{\nu},
\end{equation}

Noting that $ \widehat{a}^{\mu} = a_{\nu} \,\widehat{q}^{\mu\nu} =
a^{\mu} $ it follows that
$$\omega\equiv a^{\mu}a_{\mu} = a_{\mu} \, a_{\nu} \,\eta^{\mu\nu}
= a_{\mu} \, a_{\nu} \,\widehat{q}^{\mu\nu} = \widehat{\omega}.$$
The scalar of curvature $ \widehat{R} = \widehat{R}_{\mu\nu} \,
\widehat{q}^{\mu\nu} $ is

\begin{equation}
\widehat{R} = ( 2 + b ) \, a^{\alpha}{}_{, \alpha}.
\end{equation}

These expressions can be re-written in a covariant way by noting
that
$$ a_{\mu ; \nu} \equiv a_{\mu , \nu} -
\widehat{\Gamma}^{\epsilon}_{\mu\nu} \, a_{\epsilon}, $$ which yields

\begin{equation}
\widehat{R}_{\mu\nu} = a_{\mu \,;\, \nu} - a_{\mu} \, a_{\nu}  - (
\omega - a^{\alpha}{}_{;\,  \alpha} ) \, v_{\mu} \, v_{\nu},
\label{8jan2} \end{equation} and for the scalar $\widehat{R} $ the
form

\begin{equation}
\widehat{R} = ( 2 + b ) \, (a^{\alpha}{}_{; \alpha}- \omega ).
\end{equation}

\subsection{Analog gravity}

Suppose that an observer following a path with four-velocity $
v_{\mu}$ and acceleration $ a_{\mu}$ in the flat Minkowski
space-time background is not able to identify the origin of the
force that is acting on him. In other words he is going to believe
that only long-range gravitational forces are constraining his
motion. Let us assume that he knows that gravity does not accelerate
any curve but instead change the metric of the background according
to the principles of general relativity. This means that if he is
able to represent his motion as a geodesics in a dragged metric $
\widehat{q}_{\mu\nu} $ he will consider that the origin of such
curved metric is nothing but a consequence of a distribution of
energy which he will describe by using the equation

\begin{equation}
\widehat{R}_{\mu\nu} - \frac{1}{2} \, \widehat{R} \,
\widehat{q}_{\mu\nu} = - \widehat{T}_{\mu\nu}.
\label{8jan1}
\end{equation}

He will identify the different terms of the source through his own
motion. From his velocity $ v_{\mu} $ he defines the normalized
four-velocity $\hat{u}^{\alpha} $ in the $\widehat{Q}$ metric by
setting
$$\hat{u}^{\alpha} = \sqrt{1 + b} \, v^{\alpha}.  $$ He then
proceed to characterize the origin of the curved metric using the
standard decomposition

\begin{description}
\item
[(a)] density of energy
\begin{equation}
\widehat{\rho} = \widehat{T}_{\mu \nu} \, \hat{u}^{\mu} \,
\hat{u}^{\nu};
\end{equation}
\item
[(b)]  isotropic pressure
\begin{equation}
\widehat{p} = - \, \frac{1}{3}\, \widehat{T}_{\mu\nu} \, h^{\mu \nu};
\end{equation}
\item
[(c)] heat flux
\begin{equation}
\widehat{q}_{\lambda} = \widehat{T}_{\alpha \beta} \,
\hat{u}^{\beta} \, h^{\alpha}_{~\lambda};
\end{equation}
\item
[(d)] anisotropic pressure
\begin{equation}
\widehat{\pi}_{\mu \nu}= \widehat{T}_{\alpha \beta} \,
h^{\alpha}_{~\mu} \, h^{\beta}_{~\nu}+ \widehat{p} \, h_{\mu \nu}.
\end{equation}
\end{description}

In these expressions we used
 $$ \widehat{h}_{\mu\nu} =
\widehat{q}_{\mu\nu} - \widehat{u}_{\mu} \,\widehat{u}_{\nu}.$$ Note
that $ \widehat{h}_{\mu\nu} = h_{\mu\nu}.$ Thus, he will write
\begin{equation}
\widehat{T}_{\mu \nu}= \widehat{\rho} \, \hat{u}_{\mu} \,
\hat{u}_{\nu}- \widehat{p} \, h_{\mu \nu}+ \widehat{q}_{\mu }\
\hat{u}_{\nu}+  \widehat{q}_{\nu }\ \hat{u}_{\mu}+\widehat{\pi}_{\mu
\nu}.
\end{equation}
From this decomposition, using equation (\ref{8jan1}) and the
curvature (\ref{8jan2}) he will identify the energy-momentum
distribution as
\begin{eqnarray}
&&\widehat{\rho} =  \frac{b}{2} \, Q, \nonumber \\
&&\widehat{p} =  ( \frac{2}{3} + \frac{b}{2} ) \, Q, \label{8jan12} \\
&&\widehat{q}_{\mu} = 0, \nonumber \\
&&\widehat{\pi}_{\mu \nu} = - \, a_{\mu \, ; \, \nu} + a_{\mu} \,
a_{\nu} - \frac{Q}{3} \, \widehat{q}_{\mu\nu} + \frac{Q}{3} \,
\widehat{u}_{\mu} \, \hat{u}_{\nu }, \nonumber
\end{eqnarray}
where
$$ Q = \omega - a^{\alpha}{}_{; \, \alpha}, $$

Summarizing, we can say that this observer will state that there is
a gravitational field represented by the metric $
\widehat{q}_{\mu\nu} $ produced by the distribution of energy given
by equation (\ref{8jan12}). We note that this reduction of the
dragged metric to the framework of general relativity is not
mandatory. Indeed, we deal here precisely with some accelerated
paths that are not reduced to the gravitational force in the
standard theory. This will become more clear when we present
examples of accelerated curves in specific solutions of general
relativity in the next sections.

Indeed, let us present some clarifying examples. The first one
considers the motion of rotating bodies in Minkowski space-time. The
other ones take into account the general relativity effects. We
shall see that it is possible to produce what could be called
\textit{double gravity}, if the origin of the curvature of the
dragged metric is identified to an effective energy-momentum tensor
satisfying the equations of general relativity. However there is no
reason for this restriction. We  will come back to this question
elsewhere.

\section{Acceleration in Minkowski space-time}

Let us consider a simple example concerning the acceleration of a
body in flat Minkowski space-time written in non-stationary
cylindrical coordinate system $(t, r, \phi, z)$ to express the
following line element

\begin{equation}
\label{no_sta_met} ds^2=a^2[dt^2 - dr^2 - dz^2 + g(r)d\phi^2 +
2h(r)d\phi dt],
\end{equation}
where $a$ is a constant. We choose the following local tetrad frame
given implicitly by the $1$-forms

\begin{equation}
\begin{array}{lcl}
\theta^0&=&a(dt + h d\phi),\\[2ex]
\theta^1&=&a dr,\\[2ex]
\theta^2&=&a \Delta d\phi,\\[2ex]
\theta^3&=&a dz,
\end{array}
\end{equation}
where we define $\Delta = \sqrt{h^2 - g}$. The unique
non-identically null components of the Ricci tensor $R_{AB}$ in the
tetrad frame are

\begin{eqnarray}
R_{00}&=&\frac{1}{2a^2}\left(\frac{h'}{\Delta}\right), \nonumber \\[2ex]
R_{11}&=&\frac{1}{a^2}\left(\frac{\Delta''}{\Delta}-\frac{1}{2}\frac{h'^2}{\Delta^2}\right)=R_{22}=R_{33}, \nonumber \\[2ex]
R_{02}&=&\frac{1}{2a^2}\left(-\frac{h''}{\Delta}+\frac{h'\Delta}{\Delta^2}\right).
\label{no_sta_ricci}
\end{eqnarray}
where a prime means derivative with respect to coordinate $ r.$ The
equations of general relativity for this geometry have two simple solutions that we shall
analyze below.

In the case of $R_{AB}$=0, we get

\begin{equation}
\nonumber h'=0;\hspace{1cm} \Delta''=0.
\end{equation}
Solving these equations, we find

\begin{equation}
\nonumber h\equiv const;\hspace{1cm} \Delta\equiv\omega^2r^2,
\end{equation}
where $\omega$ is a constant. Therefore, Eq. (\ref{no_sta_met})
takes the form

\begin{equation}
 ds^2=a^2[dt^2 - dr^2 - dz^2 +
(h^2-\omega^2r^2)d\phi^2 + 2h(r)d\phi dt]. \label{no_sta_met2}
\end{equation}

If we consider the observer field

$$ v^{\mu} =\frac{1}{a\sqrt{h^2-\omega^2r^2}} \delta^{\mu}_2.$$
This path corresponds to an acceleration given by
$$ a_{\mu} = \left(0,  \frac{\omega^2r}{(h^2-\omega^2r^2)}, 0, 0 \right).$$
This means that $ a_{\mu} = \partial_{\mu} \Psi,$ where
$$ 2\Psi = - \,\ln (h^2-\omega^2r^2).$$

We are in a  situation similar to the previous section since the
acceleration is a gradient.  The parameter $ b $ of the dragged
metric is given by the expression (\ref{31dez155})

$$ 1 + b = h^2-\omega^2r^2, $$
and for the dragged metric the form

\begin{eqnarray}
\fracc{ds^2}{a^{2}} &=& \fracc{\omega^4r^4-\omega^2r^2h^2+1}{(h^2-
\omega^2r^2)^2} dt^2 + d\phi^2 \nonumber \\
&+& \fracc{2h}{h^2-\omega^2r^2}d\phi \ dt -dr^2-dz^2
\end{eqnarray}

Note that we are dealing with the case in which $ h^2-\omega^2r^2 >
0.$ This allows the presence of accelerated closed time-like curves
(CTC) in the original background that will be mapped into closed
time-like geodesics (CTG) in the dragged geometry. Note that there
exists a real singularity in $r=h/\omega$ and that $\widehat{q}_{00}
$ change sign where
$$ \omega^{2} \, r_{\pm}^{2} = \frac{h^{2} \,  \pm \, \sqrt{h^{4} -
4}}{2}.$$

\section{Acceleration in curved space-times}

Let us present now how our dragged metric approach works in some
solutions of the equations of general relativity. We choose three
well-known geometries: Schwarzschild, G\"odel universe and the Kerr
solution. In these Riemannian manifolds we analyze some examples of
accelerated paths that are interpreted as geodesics in the
associated dragged metrics.

\subsection{Schwarzschild geometry}

We set the Schwarzschild metric in the $ (t, r, \theta, \varphi) $
coordinate system
\begin{equation}
 ds^{2} = ( 1 - \frac{r_{H}}{r} ) dt^{2} - \frac{1}{( 1 -
 \frac{r_{H}}{r})} \, dr^{2} - r^{2} (d\theta^{2} + \sin^{2} \theta \, d\varphi^{2}).
\label{4dez14}
\end{equation}
Choose the path described by the four-velocity
$$ v^{\mu} = \sqrt{1 - \frac{r_{H}}{r}} \, \delta^{\mu}_{0}.$$ The
corresponding acceleration is
$$ a_{\mu} = \left( 0, - \, \frac{r_{H}}{2 (r^{2} - r \, r_{H})}, 0, 0 \right). $$
In this case the acceleration is the gradient of function $ \Psi $
given by

$$ \Psi = - \, \frac{1}{2} \,  \ln (1 - \frac{r_{H}}{r}). $$

The factor $ b $ of the dragged metric is given by
$$ b = - \, \frac{r_{H}}{r},$$ and the dragged metric takes the form

\begin{equation}
 ds^{2} =  dt^{2} - \frac{1}{( 1 -
 \frac{r_{H}}{r})} \, dr^{2} - r^{2} (d\theta^{2} + \sin^{2} \theta \, d\varphi^{2}).
\label{4dez17}
\end{equation}

The only non-null Ricci curvature of this $ \widehat{q}^{\mu\nu} $
metric are

$$R^{1}_{1} =  \, \frac{r_{H}}{r^{3}}; $$
$$ R^{2}_{2} = R^{3}_{3} = - \, \frac{1}{2} \,R^{1}_{1}. $$

All 14 Debever invariants are finite in all points except at the
origin $ r = 0.$

\subsection{G\"odel\rq s geometry}

Let us now turn our analysis to the G\"odel geometry. In the
cylindrical coordinate system this metric is given by Eq.\ (\ref{no_sta_met}),
where $a$ is a constant related to the vorticity $ a = 2/\omega^{2}$
and
$$ h(r) = \sqrt{2} \sinh^2 \ r; $$
$$ g(r) = \sinh^2 \ r (\sinh^2 \ r-1).$$
For completeness we note the non-trivial contravariant terms of the
metric: \begin{eqnarray}  g^{00} &=&   \frac{1 - \sinh^{2} r}{a^{2}
\,\cosh^{2} r}, \nonumber \\
 g^{02} &=&  \frac{\sqrt{2}}{a^{2} \,
\cosh^{2} r}, \nonumber \\
g^{22} &=&  \frac{- \, 1}{a^{2} \, \sinh^{2} r \, \cosh^{2} r}.
\end{eqnarray}

In \cite{novellonamiemilia} it was pointed out the acausal
properties of a particle moving into a circular orbit around the
$z-$axis with four-velocity
$$ v^{\mu} = \left(0, 0, \frac{1}{a \, \sinh r \, \sqrt{\sinh^{2} r
- 1}}, 0 \right). $$ This path corresponds to an acceleration given
by
 $$ a^{\mu} = \left(0,  \frac{\cosh r \, [2 \, \sinh^{2} r - 1]}{a^{2} \,  \sinh r \,
[\sinh^{2} r - 1]}, 0, 0 \right). $$ This means that $ a_{\mu} =
\partial_{\mu} \Psi,$ where
$$ \Psi = - \,\ln (\sinh r \, \sqrt{\sinh^{2}r - 1}).$$

Again, we are in a situation where the acceleration is a gradient.
Therefore, the parameter $ b $ of the dragged
metric is given by the expression (\ref{31dez155}) which in the
G\"odel\rq s background takes the form

$$ 1 + b = \sinh^{2}r ( \sinh^{2}r - 1 ), $$
and the dragged metric has the following form

\begin{eqnarray}
&&\frac{d\widehat{s}^2}{a^{2}} = \fracc{3-\sinh^4r}{(\sinh^2r-1)^2}dt^2 + d\phi^2 + 2\fracc{\sqrt{2}}{\sinh^2r-1}d\phi \ dt \ ,\nonumber\\
&&- dr^2 - dz^2
\end{eqnarray}

From the analysis of geodesics in G\"odel geometry
 the domain $ r < r_{c} $ where $
\sinh^{2}r_{c} = 1 $ separates causal from non-causal regions of the
space-time. This is related to the fact that a geodesic that pass
the value $ r = 0 $ will be confined within the domain $ \Omega_{i}
$ defined by the values of coordinate $ r $ in the region $ 0 < r <
r_{c}.$ See \cite{novellosoarestiomno} for details. However, the
gravitational field is finite in the region $ r = r_{c}.$ Nothing
similar in the dragged metric, once at $ \sinh^{2}r=1 $ there exists
a real singularity in the dragged metric. Only the exterior domain
is allowed. This means that for this kind of accelerated path in
G\"odel geometry the allowed domain for the dragged metric is
precisely the whole acausal region.

\subsection{Kerr metric}

Let us turn now to the dragged metric approach in the case the
background is the Kerr metric. In the Boyer-Lindquist coordinate
system this metric is given by

\begin{equation}
\begin{array}{l}
\label{ds2kerr}
ds^2=\left(1-\fracc{2Mr}{\rho^2}\right)dt^2-\fracc{\rho^2}{\Sigma}dr^2-\rho^2d\theta^2+\\[2ex]
+\fracc{4Mra\sin^2\theta}{\rho^2}dtd\phi+\\[2ex]
-\left[(r^2+a^2)\sin^2\theta+\fracc{2Mra^2\sin^4\theta}{\rho^2}\right]d\phi^2,
\end{array}
\end{equation}
where $\Sigma=r^2+a^2-2Mr$ and $\rho^2 = r^2+a^2\cos^2\theta$. On
equatorial plane ($\theta=\pi/2$) consider the following vector
field

$$ v^{\mu} = \left(0, 0, 0, \fracc{r}{\sqrt{-(r^2+a^2)^2+a^2\Sigma}} \right). $$
This path corresponds to an acceleration given by

$$ a_{\mu} = \left(0, -\fracc{r^3-Ma^2}{r^4+r^2a^2+2Mra^2}, 0, 0 \right).
$$ This means that $ a_{\mu} = \partial_{\mu} \Psi,$ where

$$ 2\Psi = -\ln\left[-\left(r^2+a^2+\fracc{2Ma^2}{r}\right)\right].$$

Once more we choose an accelerated  path that can be represented by a
gradient. The parameter $ b $ of the dragged metric is given by the
expression (\ref{31dez155})

$$ 1 + b = -\left(r^2+a^2+\fracc{2Ma^2}{r}\right),$$
and for the dragged metric, on the equatorial plane, the form

\begin{equation}
\begin{array}{l}
\fracc{ds^2}{a^{2}}= \fracc{1}{(1+b)^2}\left(1-\fracc{2M}{r}-b\fracc{4M^2a^2}{r^2}\right)dt^2 +d\phi^2\\[2ex]
+ \fracc{4Ma}{(1+b)r}dtd\phi - \fracc{r^2}{\Delta}dr^2,
\end{array}
\end{equation}

These two last cases (G\"odel and Kerr) show a very
curious and intriguing property: the accelerated CTC's
at their respective metrics) are transformed in curves that are
geodesics, that is CTG's (at their corresponding dragged metric).
Besides, in both cases, the dragged metrics display a real
singularity excluding the causal domain.
\section{General case}

In the precedent sections we limited our analysis to the case in
which the acceleration is given by a unique function. Let us now
pass to more general situation. In order to geometrize any kind of
force we must deal with a larger class of geometries. The most
general form of dragged metric that allows the description of
accelerated bodies as true geodesics in a modified geometry has the
form
 \begin{equation}
\widehat{q}^{\mu\nu} =  g^{\mu\nu} + b \, v^{\mu} \, v^{\nu} + m \,
a^{\mu} \, a^{\nu} + n ( v^{\mu} \, a^{\nu} + a^{\mu} \, v^{\nu}).
\label{301216}
\end{equation}
The three parameters $ b, m, n$ are related to the three degrees of
freedom of the acceleration vector. The corresponding covariant form
of the metric is given by

 \begin{equation}
\widehat{q}_{\mu\nu} =  g_{\mu\nu} + B \, v_{\mu} \, v_{\nu} + M \,
a_{\mu} \, a_{\nu} + N ( v_{\mu} \, a_{\nu} + a_{\mu} \, v_{\nu}).
\label{30121620}
\end{equation}
in which $ B, M, N$ are given in terms of the parameters $ b, m,n$
by the relations
$$ B = - \, \frac{b \, ( 1 + m \, \omega ) - n^{2} \,\omega}{(1 +
b) \, ( 1 + m \, \omega) - n^{2} \, \omega}, $$

$$ M = \frac{1}{1 + m\, \omega} \, \left( - \, m +
\frac{n^{2}}{(1 + b) \, (1 + m \,\omega) - n^{2} \, \omega}
\right),$$

$$ N = - \, \frac{n}{ (1 + b) \, (1 + m \,\omega) - n^{2} \, \omega}.
$$

In this case the equation that generalizes the condition of
geodesics (\ref{31dez1}) has the form
\begin{eqnarray}
a_{\mu} &=& \frac{1}{2} \, \left( (1 + b) \, v^{\lambda} \, v^{\nu}
+ n \, a^{\lambda} \, a^{\nu}\right)  \, [\widehat{q}_{\lambda\mu \,
, \, \nu}   \nonumber \\
&+&  \widehat{q}_{\lambda\nu \, , \, \mu} - \widehat{q}_{\mu \nu \,
, \, \lambda}].
\end{eqnarray}
This equation can be cast in the following formal expression
\begin{equation}
a_{\mu} = \Psi_{1} \, \partial_{\mu} b + \Psi_{2} \, \partial_{\mu}
m +  \Psi_{3} \, \partial_{\mu} n ,
\end{equation}
where each term  $\Psi_{1},  \Psi_{2}$ and $\Psi_{3}$  depends on
all three functions $ m, b $ and $ n .$

Solving this equation for these three functions provide the most
general expression for any acceleration.

With these results, we have transformed the path of any particle
submitted to any kind of force as a geodetic motion in the dragged
metric. This result is the extension of the d'Alembert principle,
corresponding to all types of motion i.e. the acceleration is
geometrized through the dragged metric.

\section{Conclusion}

We summarize the novelty of our analysis in the following steps:

\begin{itemize}
\item{Let $v_{\mu}$ represent the four-vector that describes the kinematics of a body in an arbitrary
space-time endowed with a geometry $ g_{\mu\nu};$}
\item{If the body experiences a non-gravitational force it acquires an acceleration
$a_{\mu}$;}
\item{It is always possible to define an associated dragged metric
 $ \widehat{q}_{\mu\nu}$ given by (\ref{30121620}) such that in this metric the
acceleration is removed. That is, the path is represented as a free
particle that follows geodesics in the dragged geometry.}
\end{itemize}

We have shown by a constructive operation that the following
conjecture is true:

\begin{theorem}
For any accelerated path $ \Gamma$ described by four-vector velocity
$ v_{\mu}$ and acceleration $ a_{\mu}$ in a given Riemannian
geometry $ g_{\mu\nu} $ we can always construct another geometry $
\widehat{Q}$ endowed with a dragged metric $ \widehat{q}_{\mu\nu} $
which depends only on $ g_{\mu\nu},  v_{\mu}$ and $ a_{\mu}$ such
that the path $ \Gamma $ is a geodesic in $ \widehat{Q}.$
\end{theorem}

\section{acknowledgements}
We would like to thank Dr Ivano Dami\~ao Soares for his comments in
a previous version of this paper. We would like to thank FINEP, CNPq
and FAPERJ and EB CNPq for their financial support.

\end{document}